\documentclass[a4paper,fleqn]{cas-dc}

\usepackage[numbers]{natbib}

\def\tsc#1{\csdef{#1}{\textsc{\lowercase{#1}}\xspace}}
\tsc{WGM}
\tsc{QE}
\tsc{EP}
\tsc{PMS}
\tsc{BEC}
\tsc{DE}

\begin{document}
\let\WriteBookmarks\relax
\def\floatpagepagefraction{1}
\def\textpagefraction{.001}
\shorttitle{Automatic Pulmonary Artery-Vein Separation in CT Images using Twin-Pipe Network and Topology Reconstruction}
\shortauthors{L. Pan, Y. Zheng et al.}

\title [mode = title]{Automatic Pulmonary Artery-Vein Separation in CT Images using Twin-Pipe Network and Topology Reconstruction}                      
\tnotemark[1]

\tnotetext[1]{This work was supported in part by the Natural Science Foundation of Fujian Province, Grant/Award Number: 2020J01472; in part by the Provincial Science and Technology Leading Project, Grant/Award Number: 2018Y0032.}

\author[1]{Lin Pan}
\author[1]{Yaoyong Zheng}
\author[1]{Liqin Huang}
\author[1]{Liuqing Chen}
\author[1]{Zhen Zhang}
\author[2]{Rongda Fu}
\author[3]{Bin Zheng}

\author[1]{Shaohua Zheng}[orcid=0000-0002-9133-705X]
\cormark[1]

\address[1]{School of Physics and Information Engineering, Fuzhou University, Fuzhou 350108, China}
\address[2]{ School of Mechanical engineering and Automation, Fuzhou University, Fuzhou 350108, China}
\address[3]{the thoracic department, Fujian Medical University Union Hospital, Fuzhou, China}

\cortext[cor1]{Corresponding author}
\ead{sunphen@fzu.edu.cn}

\begin{abstract}
With the development of medical computer-aided diagnostic systems, pulmonary artery-vein(A/V) separation plays a crucial role in assisting doctors in preoperative planning for lung cancer surgery. However, distinguishing arterial from venous irrigation in chest CT images remains a challenge due to the similarity and complex structure of the arteries and veins. We propose a novel method for automatic separation of pulmonary arteries and veins from chest CT images. The method consists of three parts. First, global connection information and local feature information are used to construct a complete topological tree and ensure the continuity of vessel reconstruction. Second, the Twin-Pipe network proposed can automatically learn the differences between arteries and veins at different levels to reduce classification errors caused by changes in terminal vessel characteristics. Finally, the topology optimizer considers interbranch and intrabranch topological relationships to maintain spatial consistency to avoid the misclassification of A/V irrigations. We validate the performance of the method on chest CT images. Compared with manual classification, the proposed method achieves an average accuracy of 96.2\% on noncontrast chest CT. In addition, the method has been proven to have good generalization, that is, the accuracies of 93.8\% and 94.8\% are obtained for CT scans from other devices and other modes, respectively. The result of pulmonary artery-vein obtained by the proposed method can provide better assistance for preoperative planning of lung cancer surgery.
\end{abstract}


\maketitle

\section{Introduction}
With the development of computed tomography (CT) technology, multislice CT image has become the main auxiliary tool for thoracic surgeons to screen, diagnose, and treat diseases and prognosis. Pulmonary diseases are often accompanied by pulmonary A/V lesions. In recent years, pulmonary vessel segmentation has been widely explored, but the potential relationship between pulmonary diseases and pulmonary A/V is unknown. Pulmonary vessel segmentation has been unable to satisfy the clinical requirements\cite{zheng2019report}.
\begin{figure}[t]
\centerline{\includegraphics[width=\columnwidth]{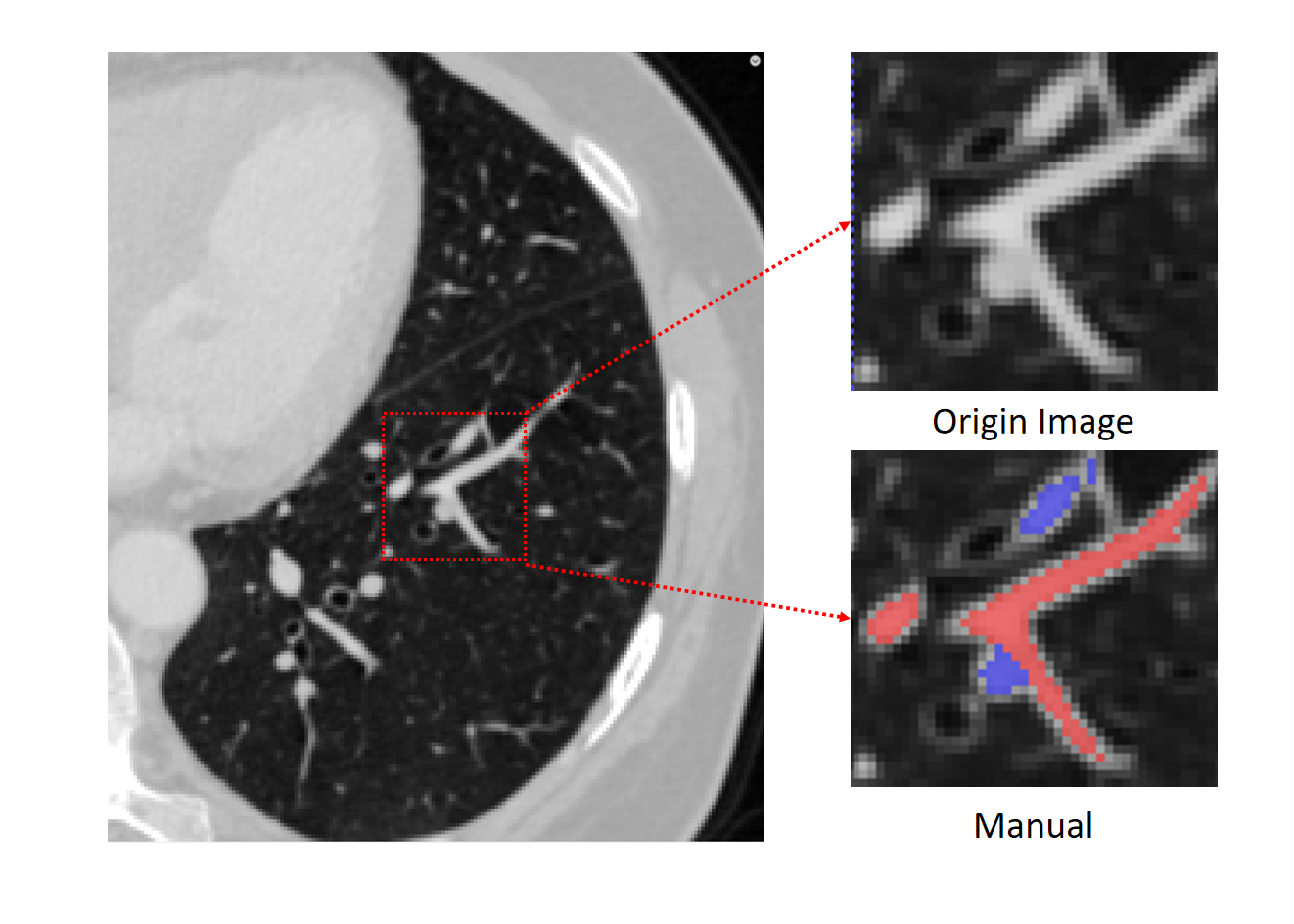}}
\caption{In an example of a pulmonary CT image that is noncontrast and a doctor's manual standard, arteries (blue areas) and veins (red areas) are close to each other and share similar strength values.}
\label{fig1}
\end{figure}

Artery-vein segmentation is useful for various diseases or pathological conditions. Surgeons can reconstruct pulmonary arteries and veins through medical computer-aided diagnosis systems and analyze the anatomical relationship between pulmonary disease and the vascular system, such as vascular geometric shape, topological structure, spatial scale, and other characteristics. Then, the location of the lesion is accurately located, and the surgical plan is formulated to improve the success rate of the surgery. For example, the growth of lung tumors may affect the surrounding vascular system as well as the bronchi. Diagnosis of lung cancer requires differentiation of bronchi, pulmonary A/V, including determination of their relative spatial relationship to suspected pulmonary nodules because of the complex structure of these organs\cite{tozaki2001extraction}. However, due to the large amount of CT data, manual analysis of this process is more time-consuming and visually demanding, which greatly increases the burden on doctors. Therefore, the automatic separation of pulmonary artery-vein is greatly important for clinical diagnosis.

At present, the automatic separation of pulmonary arteries and veins from multislice CT is always an open problem, especially in noncontrast CT scans due to the following reasons: 1) A/V is indistinguishable by its similar intensity values in noncontrast CT images. 2) The vascular tree structure is extremely complex and dense, with arteries and veins close to each other and intertwined. 3) Artifacts, partial volume effects, and patient-specific vascular tree structural abnormalities cause difficulty of A/V segmentation. CT provides images with near-isotropic submillimeter resolution. It enables a detailed display of the pulmonary vessels. The details are shown in Fig. 1. In addition, it can be seen that it is difficult for doctors to make an accurate manual standard based on voxel levels from the original CT images.

\subsection{Background and state-of-the-art approach on A/V segmentation}

\begin{figure}[t]
\centerline{\includegraphics[width=\columnwidth]{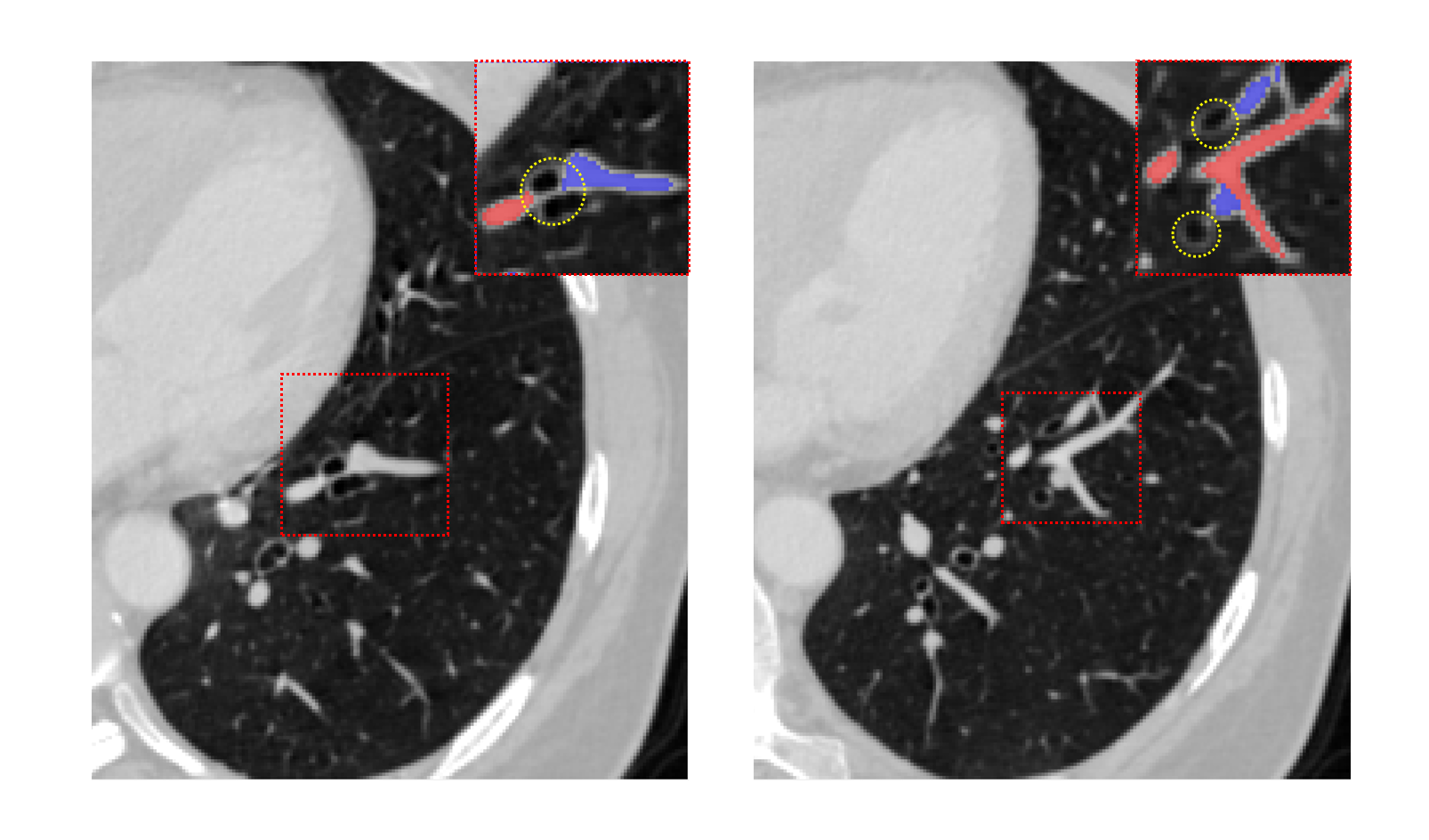}}
\caption{In a case of CT, the process of pulmonary vessels moving away from the hilum of the lung and accompanied by bronchi is described. Vessels near the hilum (the left column) and vessels away from the hilum (the right column), the red dashed rectangles highlights the airway (yellow dotted circle) is associated with arteries and veins at different levels.}
\label{fig2}
\end{figure}
With the development of AI-assisted medical imaging diagnostic system, many scholars have tried to propose their own pulmonary A/V separation methods in the last decade, although it is very difficult. Despite the vasculature inside the lungs is very variable, some inherent anatomical properties are usually found. One property is that arteries run along the bronchi, and the veins run in between their branches. As shown in Fig. 2, the trachea with the accompanying artery is not evident when the vessels are near the hilum of the lung. As the artery moves away from the hilum, the bronchi begin to follow the artery closely. Some methods rely on the bronchi for A/V separation. Tozaki et al.\cite{tozaki2001extraction} used information about distances between vessel segments and bronchi to separate the arteries and veins. Y.Mekada et al.\cite{nakamura2005pulmonary} classified pulmonary artery and vein from X-ray CT images according to the distance from the bronchus region to the vessel segment and the distance between the nearest inter lobar to the vessel. Similarly, Buelow et al.\cite{bulow2005automatic} designed a method of “arterialness” to classify each extracted vessel segment. However, these methods of A/V separation depend on the quality of airway segmentation.

Another type of A/V separation studies focuses more on the vascular spatial structure. Saha et al.\cite{saha2010topomorphologic} found the size of the local kernel according to the size of the vessels and tracked the continuity of the local separation area to separate the location of attached arteries and veins. Wala et al.\cite{wala2011automated} designed an automated trace-based separation scheme from low-dose CT scans that tracked arterial seed point from the basal pulmonary artery region and detected bifurcation to separate the artery from other nearby isometric structures. Kitamura et al.\cite{kitamura2016data} used vascular connection information to design a method based on high-order potentials energy minimization to classify pulmonary vascular voxels. Park et al.\cite{park2013automatic} proposed a method to find voxel-based trees with minimal construction energy inside the vessel segmentation and divided them into a group of subtrees that share the same AV classification. These methods often require manual intervention or depend on specific CT quality.

In recent years, several new ideas for automatic A/V separation have been proposed. Payer et al.\cite{payer2015automatic} combined their previous ideas and proposed a fully automated algorithm for arterial and venous separation in thoracic CT images based on two integer programming. Through integer programming, parameter tuning was carried out to extract subtrees, and the A/V separation was performed. Charbonnier et al.\cite{charbonnier2015automatic} proposed a method to convert the segmentation of vessels into a geometric graph representation. The method used local information to detect the attached arteries and veins in the geometric graph to generate subgraphs. Then, the volume difference between arteries and veins was used for classification. Daniel et al.\cite{jimenez2019graph} proposed a new scheme for the separation of pulmonary A/V trees by segmenting the tree-like structure and designing a specialized graph-cut method for classification to ensure the connectivity and consistency of the vascular derived subtrees. However, these methods require parameter tuning and are parameter sensitive.

Many studies attempt to solve the problem of A/V separation with deep learning due to the better robustness and stability of deep network. Nardelli et al.\cite{nardelli2018pulmonary} proposed a convolutional neural network for A/V separation of pulmonary lobar vessel particles. The final classification result was obtained by graph-cut strategy that combined connectivity and preclassification information. Zhai et al.\cite{zhai2019linking} represented vascular tree as a graph based on vessel segmentation and skeletonization. Then, a CNN-GCN classifier was designed to combine local image information with graph connection information to train the constructed container graph, and the A/V separation results were obtained. Yulei Qin et al.\cite{qin2020learning} proposed a learning tubular-sensitive CNNs for pulmonary A/V classification, which had superior sensitivity to arterioles and venules.

Although these studies have certain theoretical basis and achieved considerable results, the results after separation still have some problems, such as A/V discontinuity or A/V confusion. Thus, the approach still has a certain distance from clinical application.

\subsection{Our contribution}
To address these concerns, we propose pulmonary A/V separation based on Twin-Pipe network in noncontrast CT images. First, a vascular tree topology structure, which flexibly combine a scale-space particle algorithm and the multistencil fast marching algorithm(MSFM), is constructed to ensure the continuity and authenticity of topology reconstruction. Second, a Twin-Pipe classification network is designed, which can effectively learn the location characteristics of different vessels, including the trachea closely associated with the artery at the terminal vessel. Finally, we consider interbranch and intrabranch topological relationships and propose a pulmonary A/V topology optimizer to avoid the misclassification caused by A/V irrigations.

The method is evaluated by the accuracy of vessel particles in noncontrast CT cases. In addition, the generalization of the method under different CT imaging devices and different modes is fully verified. To perform an appropriate comparison, the method is also evaluated on annotated CT data in CARVE14 dataset\cite{charbonnier2015automatic}.

The rest of this paper is shown as follows. we propose an approach to solve the A/V separation problem in noncontrast CT images in Section 2, which includes vascular tree topology extraction, Twin-Pipe network preliminary classification, and pulmonary A/V topology optimizer to refine the classification results. Section 3 mainly introduces the experiment methods, including data sources, experimental setup details, and evaluation metrics. Then, in Section 4, the experimental results of this method, ablation experiment, and generalization experiment are presented. Finally, Sections 5 elaborates the discussion and conclusion of this paper.

\begin{figure*}
\centerline{\includegraphics[width=1\linewidth]{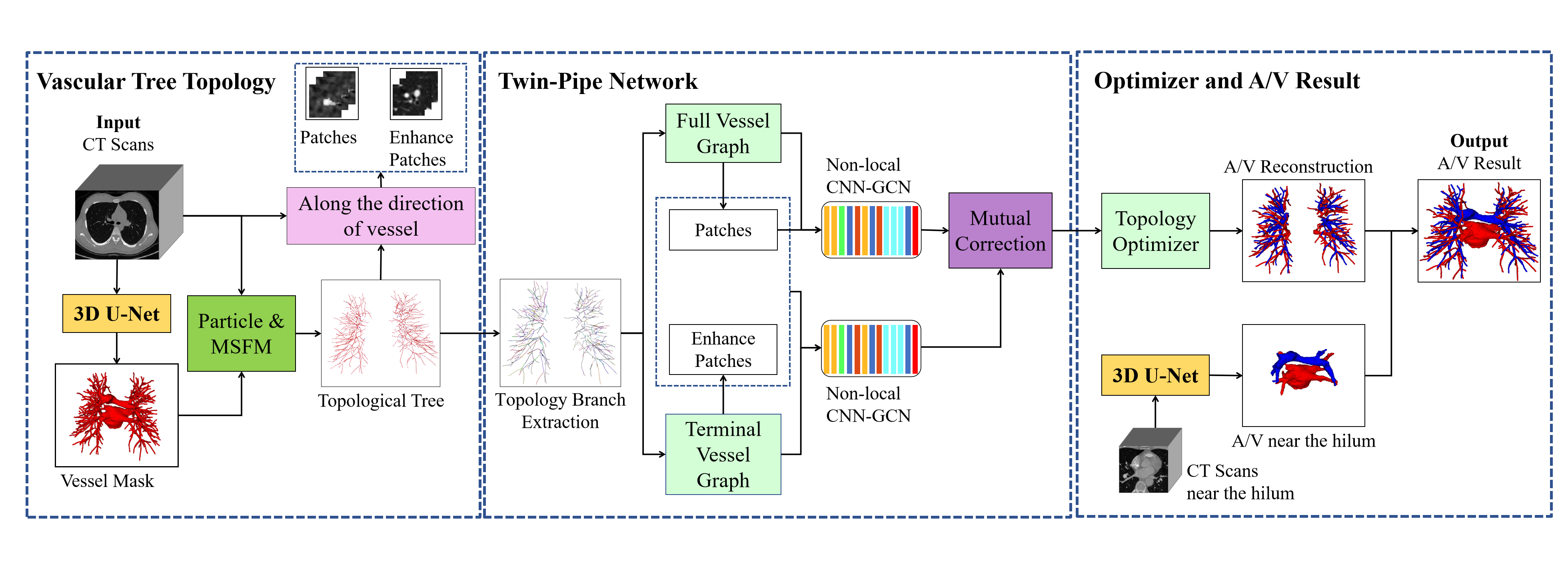}}
\caption{Overview of the proposed pulmonary A/V separation method. It includes vascular tree topology extraction, Twin-Pipe network, topology optimizer and A/V results.}
\label{fig3}
\end{figure*}
\begin{figure}[t]
\centerline{\includegraphics[width=\columnwidth]{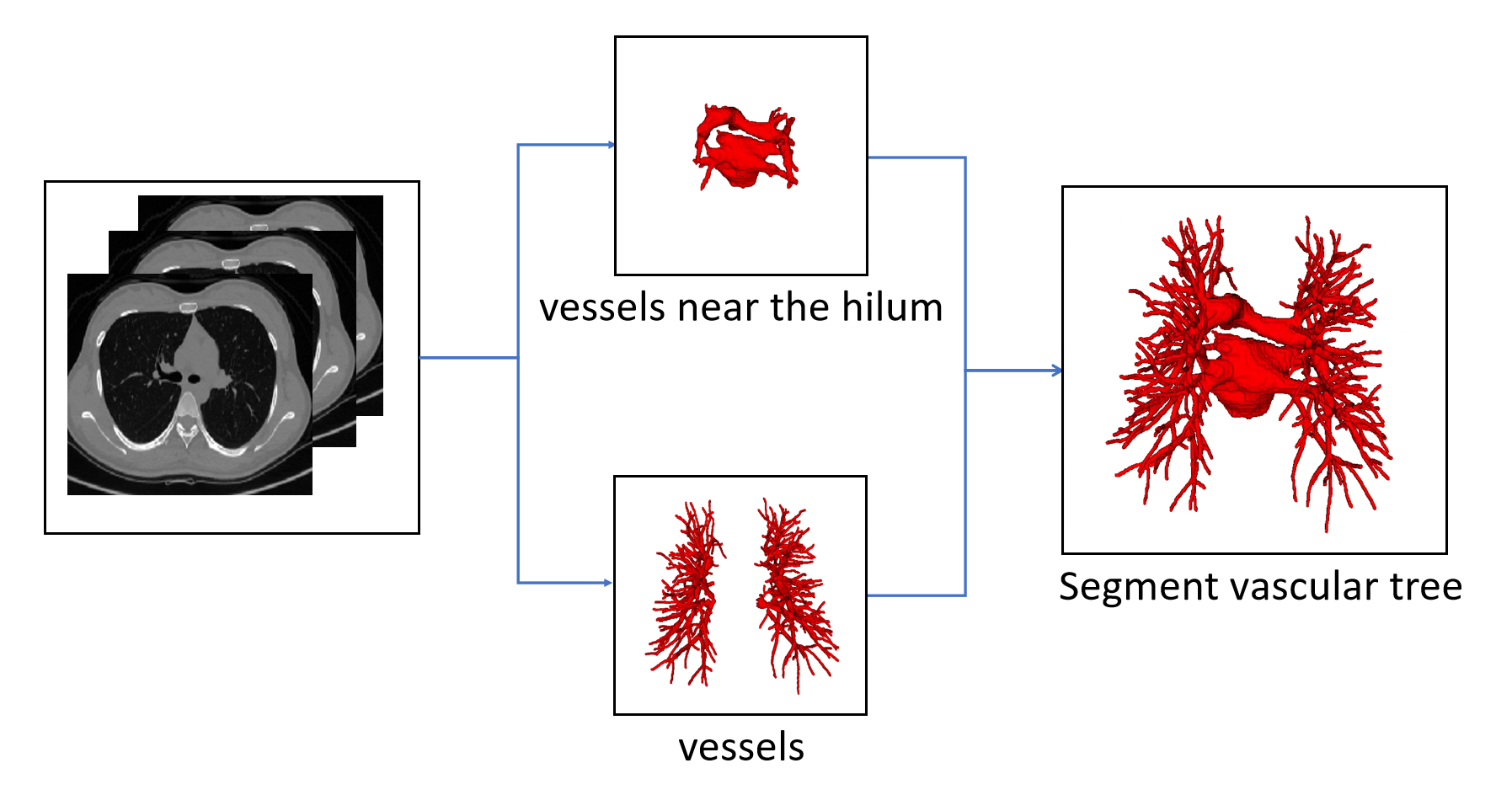}}
\caption{An overview of the proposed pulmonary vascular tree segmentation process. The vessels and vessels near the hilum are segmented and fused from the original CT images.}
\label{fig4}
\end{figure}

\section{Methods}

The overall framework of pulmonary A/V separation method in this paper is shown in Fig. 3, including vascular tree topology extraction, preliminary classification of Twin-Pipe network and topology optimizer refined classification pulmonary A/V results.

In the vascular tree topology extraction module, the vessels and the vessels near the hilum of the lung are segmented from thoracic CT images respectively, and a full vascular tree can be obtained by fusing them together. Then the topology is extracted, and the distance transform is used to guide and compensate the missing points flexibly. The vascular tree is represented as a collection of particles, and the vessel particles are transformed into topological trees. In Twin-Pipe network, 3D patches are taken from each particle on the topological tree as the center, and the whole vessel segment and the terminal branch vessel are trained separately from the topological tree, and the preliminary classification results of vessel particles are obtained. Finally, in optimizer and A/V result module, topological subtrees and topological branches are extracted from the topological tree, and the branch confidence is calculated to prune the subtree. Optimization of A/V classification using topological connectivity. Then, the pulmonary A/V topology is reconstructed and the A/V near the hilum are fused to obtain the pulmonary A/V results.

\subsection{Vascular tree topology}

\subsubsection{Vascular tree segmentation}
{The proposed method begin with vascular tree segmentation. Vascular trees are extracted from chest CT scan by fusion of vessels\cite{cui2019pulmonary} and vessels near the hilum of the lung. The specific process is shown in Fig. 4. And then, topology tree is constructed by skeleton topology extraction.}

\begin{figure}[t]
\centerline{\includegraphics[width=\columnwidth]{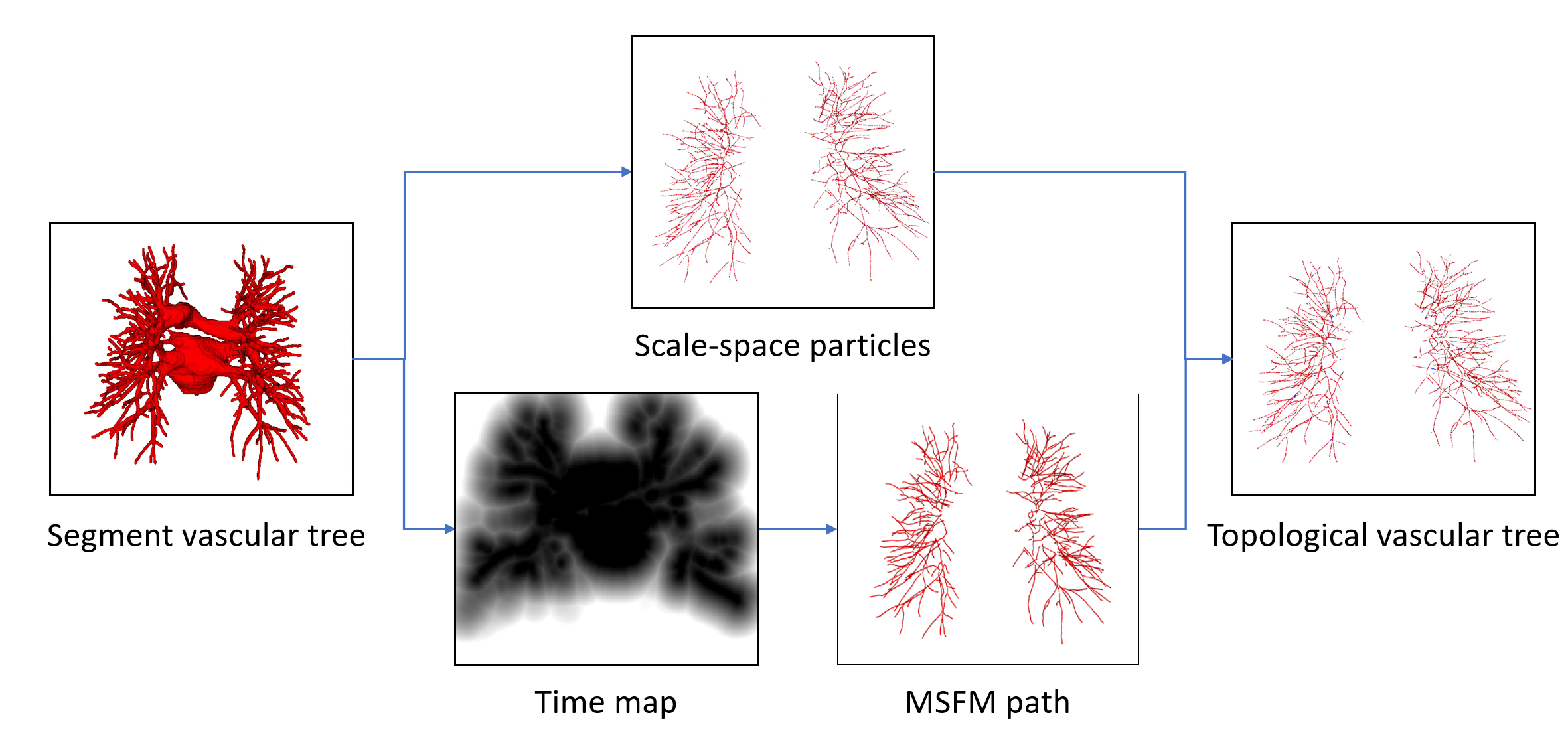}}
\caption{An overview of the proposed topological vascular tree extract process. The initial topology is obtained by the scale-space particles, and the final topological vascular tree is obtained by refining the initial topology with the guidance of time map.}
\label{fig5}
\end{figure}

\subsubsection{Vascular tree topology extraction}
{Due to the applicability of the skeleton algorithm and the characteristics of the vascular tree itself, it is often impossible to obtain accurate results of skeleton structure\cite{saha2016survey}. we flexibly combine the Multi-Stencils Fast Marching\cite{liu2018automated} and a scale-space particles reconstruction algorithm\cite{estepar2012computational} to develop a new topomorphologic algorithm  to construct the topological vascular tree.

Firstly, based on the characteristics of vascular tubular structure, the scale-space particle reconstruction algorithm was used to extract the initial skeleton of the vessel and obtain the local information of the vessel mask. Scale-space particles sampling method exploits the theory of linear scale-space to localize the features of the image described by the Hessian. The vascular tree is represented as a set of particles that each contains vessel scale, orientation, and intensity information. Therefore, particles are represented as X = \{$x_i$\}, where $x_i$ represents a particle point. However, after vascular tree reconstruction, the vessels are discontinuous, which is due to the inability of this method to identify non-tubular structures, such as the junction of vessels, resulting in partial point loss. At the same time, there was no parent-child relationship between the particles.

Secondly, the skeleton extraction algorithm based on distance transform can maintain the connectivity of the vascular tree. MSFM is used to obtain the global information from the vessel mask, and the vessel point path is obtained by iteratively tracing from the end of the potential tree to the root node, where the branch online confidence score is calculated in the time map to determine whether the trace iteration should be updated or excluded from the skeleton tree. In this paper, 3D vessel mask is taken as the input with 3D coordinates x, aiming to output the vessel skeleton tree G, and a 3D spatial coordinate and radius are specified for each vessel skeleton tree G node. The degree of each vascular tree node can be between 1 and 3. However, the radius of each node of the obtained vessel tree is only an approximation and does not really reflect the vascular tubular shape, and the results after reconstruction are different from the real vascular tree to some extent.

\begin{figure}[t]
\centerline{\includegraphics[width=\columnwidth]{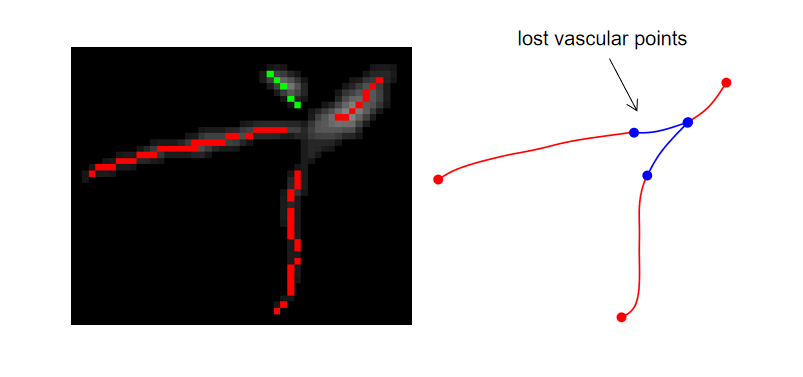}}
\caption{Taking the relationship between 2D distance map and topological trajectory as an example, skeleton points were missing at vessel bifurcation, and the shortest path is calculated by distance map through the false positive terminal point position information to make up for the missing skeleton. In the same branching path, The red represents the extracted skeleton and the blue represents the missing skeleton. The green belongs to the skeleton extracted by another branch.}
\label{fig6}
\end{figure}

Finally, in order to ensure the continuity and authenticity of the topology reconstruction, we flexibly combined the global and local information of the vascular skeleton. In this paper, a particle-based 26-neighborhood search method is used to extract the topological structure of the final vessels by guiding the time map to make up for the missing vessel particles. In this paper, vessel particles can be classified into three categories: terminal points (or false-positive terminal points), branching points, and bifurcating points. We indicate the number of points in the 26 neighborhoods of vessel particles $x_i$ as $\Omega_{26}$($x_i$). When the particle $x_i$ is the terminal point, $\Omega_{26}$($x_i$) = 1; when the particle $x_i$ is the branching points, $\Omega_{26}$($x_i$) = 2; when the particle point $x_i$ is the bifurcating points, $\Omega_{26}$($x_i$) \textgreater 2. Due to the presence of branch point loss or branch fracture resulting in false positive terminal points, further discrimination is needed. Travel in the direction of the terminal vessel points, and if the terminal point is still on the vessel mask after traveling, it is considered as a false positive terminal point, where the travel distance is between one and two scales. Otherwise, it is considered to be the true terminal point. For the false positive terminal points, the trajectory of the lost vessel points is obtained by MSFM in the time map, in which the time map is calculated from the 3D distance map. Finally, the complete topology tree is obtained. The process of the vascular tree topology extraction method is shown in Fig. 5 and Fig. 6. And compared with the scale-space particle reconstruction algorithm, the advantages of the vascular tree topology extraction in this paper are shown in Fig. 7.}

\begin{figure*}
\centerline{\includegraphics[width=1\linewidth]{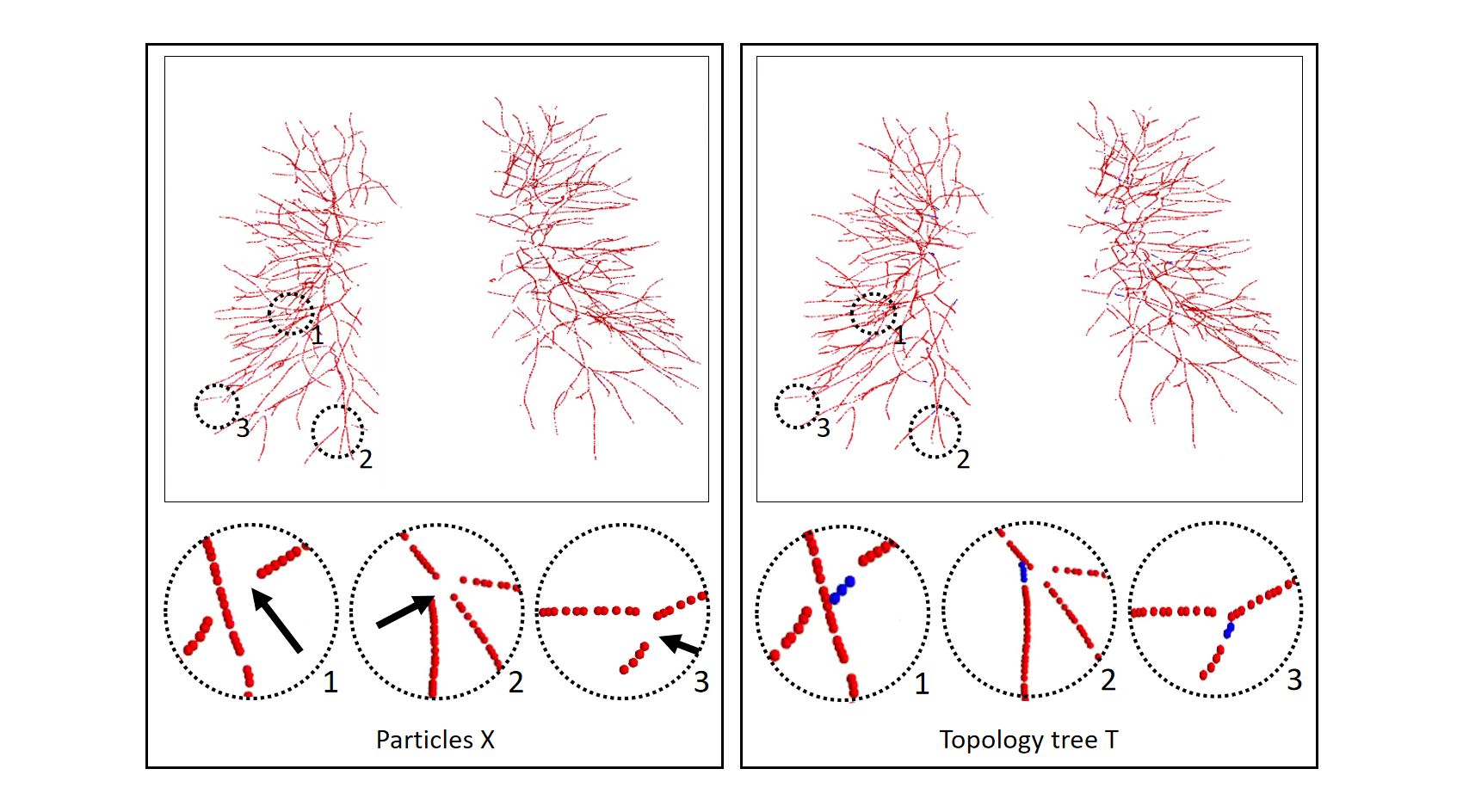}}
\caption{An example result of a vascular tree topology extraction method. The left column is the result of the scale-space particles, and the right column is the result of our proposed vascular tree topology extraction, with the circle highlighting the local skeleton points extracted from the topology. Due to the complex and changeable structure of the vascular tree, the local circular display position has been translated, rotated and enlarged from the original image position, which is different from the original image position to some extent. The black arrow points to the area where the topology points are lost, and the blue topology points indicate that our method makes up the missing points.}
\label{fig7}
\end{figure*}

\subsection{Preliminary classification of Twin-Pipe network}

\subsubsection{A Non-local CNN-GCN classifier}
{Traditional CNN network has limitation in capturing long range dependencies which extract the global understanding of visual scenes. In this paper, we consider both image information and connectivity information by connecting Non-local CNN and GCN. The Non-local CNN considers the global context information of the image, while the GCN module can learn the local and graph connection information. Combining these two modules together is useful for analyzing the vascular tree. In order to connect the Non-local CNN network with the GCN network, this paper adapt the end-to-end method of CNN connecting GCN proposed by Zhai et al.\cite{zhai2019linking}. In the preliminary classification of A/V, the learning ability of the Non-local CNN-GCN method for the A/V features is better than that of the traditional CNN-GCN, Non-local CNN3D and CNN3D.

\begin{figure}[t]
\centerline{\includegraphics[width=\columnwidth]{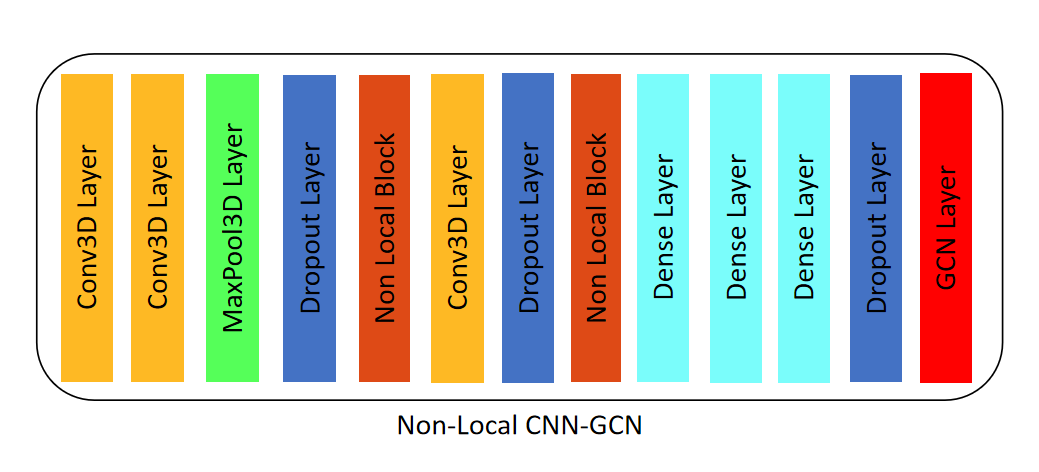}}
\caption{Architecture of the Non-local CNN-GCN classifier.}
\label{fig8}
\end{figure}

In this paper, for each point on the vascular topology tree, the orientation information of each node is used to extract the local patch of size S = [32, 32, 3] that should be perpendicular to the vascular direction from the CT image. A patch is labeled either artery or vein,i.e. $y_i$  $\in$ \{1, 2\}, based on the label of the center voxel. We randomly select node b = 128 and their patches from our graph structure T as input, and we need the image patches of these neighbors n. Take NB of size b × n × s as the input to the Non-local CNN-GCN network, where n is the number of neighbors. Then the Non-local CNN-GCN classifier presented in this paper is trained and each center voxel is predicted to be an artery or vein. The Non-local CNN-GCN classifier adopted in this paper is shown in Fig. 8.}

\begin{figure*}
\centerline{\includegraphics[width=1\linewidth]{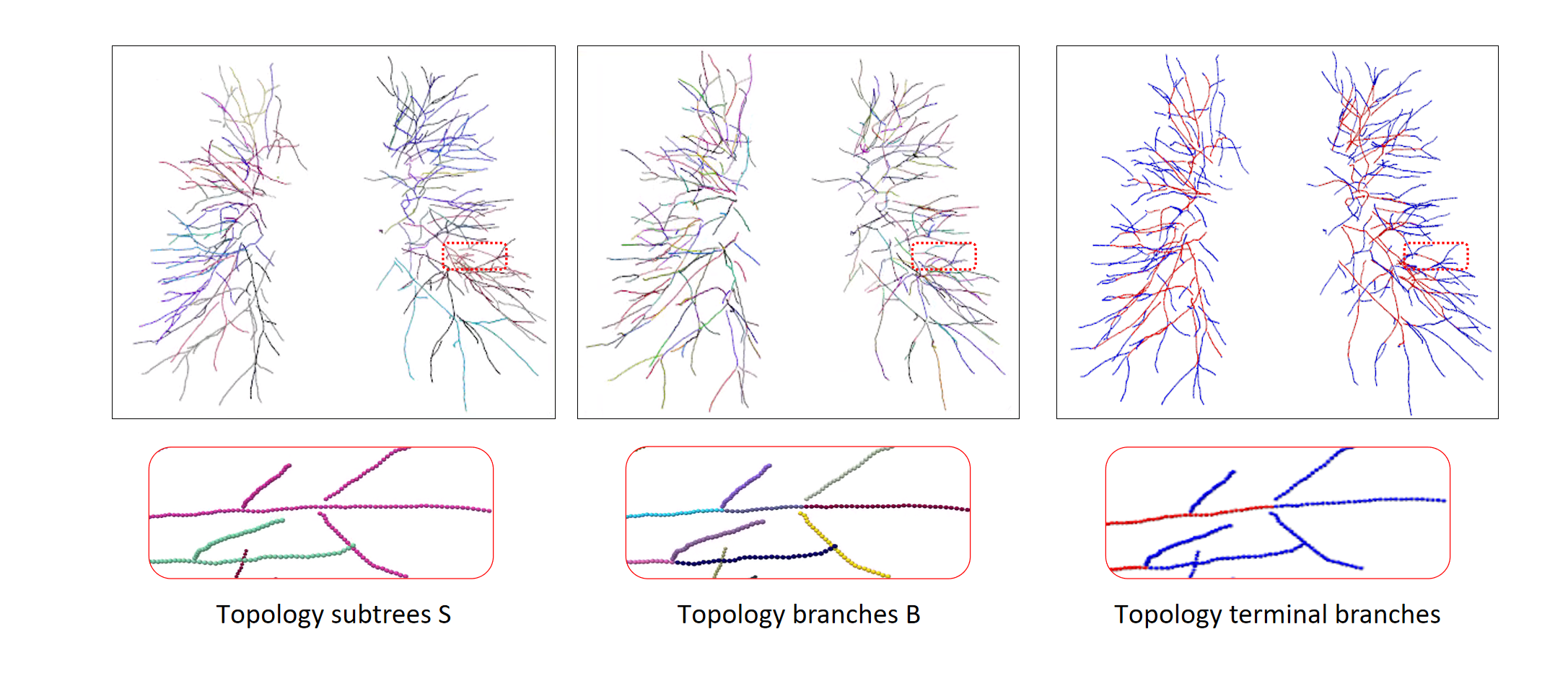}}
\caption{An example of topology terminal branch extraction process, from left to right column are topology subtree, topology branch, topology terminal branch. The different color segments represent different categories. Due to the complex and changeable structure of the vascular tree, the local red frame display position has been translated, rotated and enlarged from the original image position, which is different from the original image position to some extent.}
\label{fig7}
\end{figure*}

\begin{figure}[h]
\centerline{\includegraphics[width=\columnwidth]{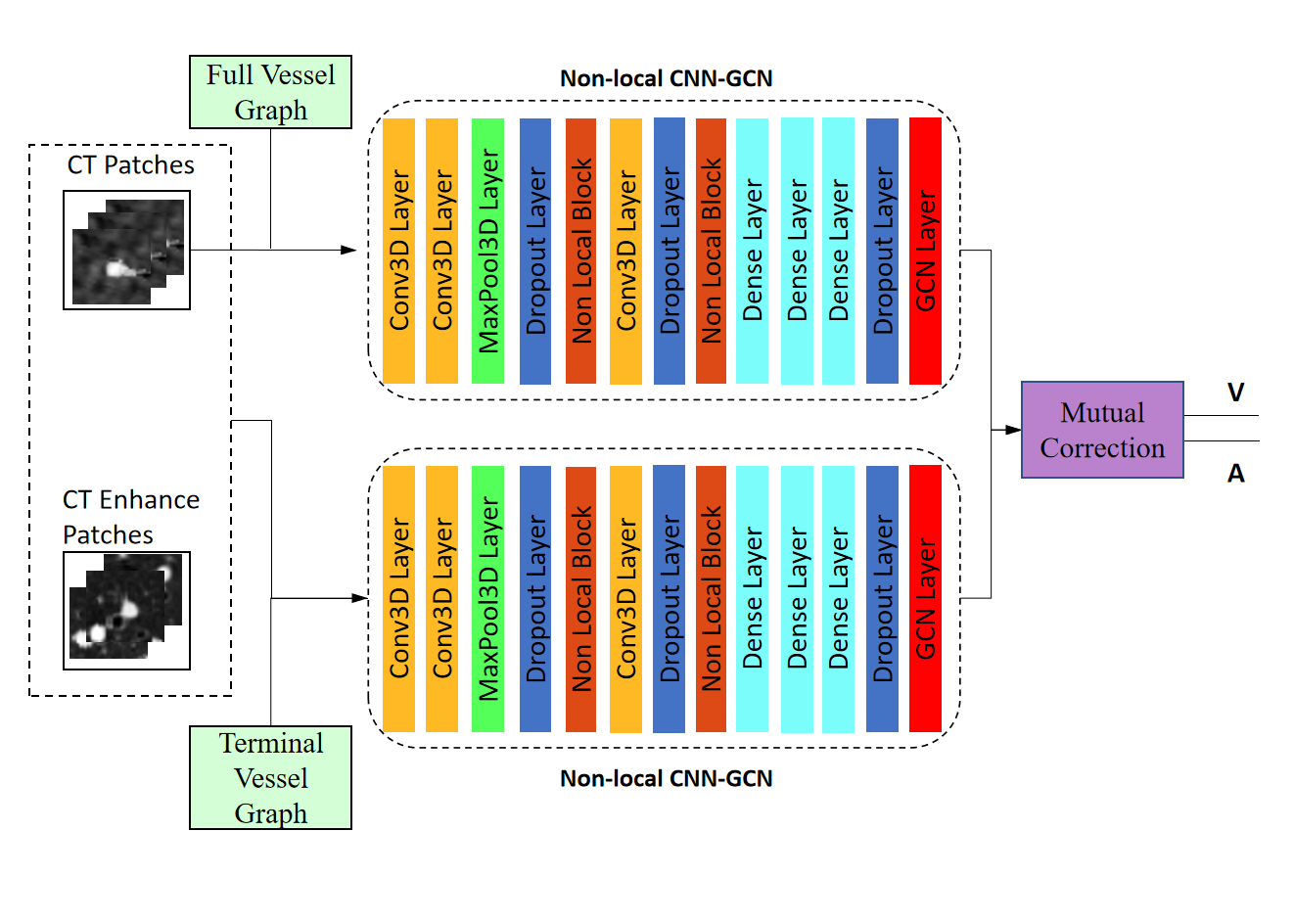}}
\caption{Architecture of the Twin-Pipe network for preliminary A/V separation.}
\label{fig10}
\end{figure}

\begin{figure*}
\centerline{\includegraphics[width=1\linewidth]{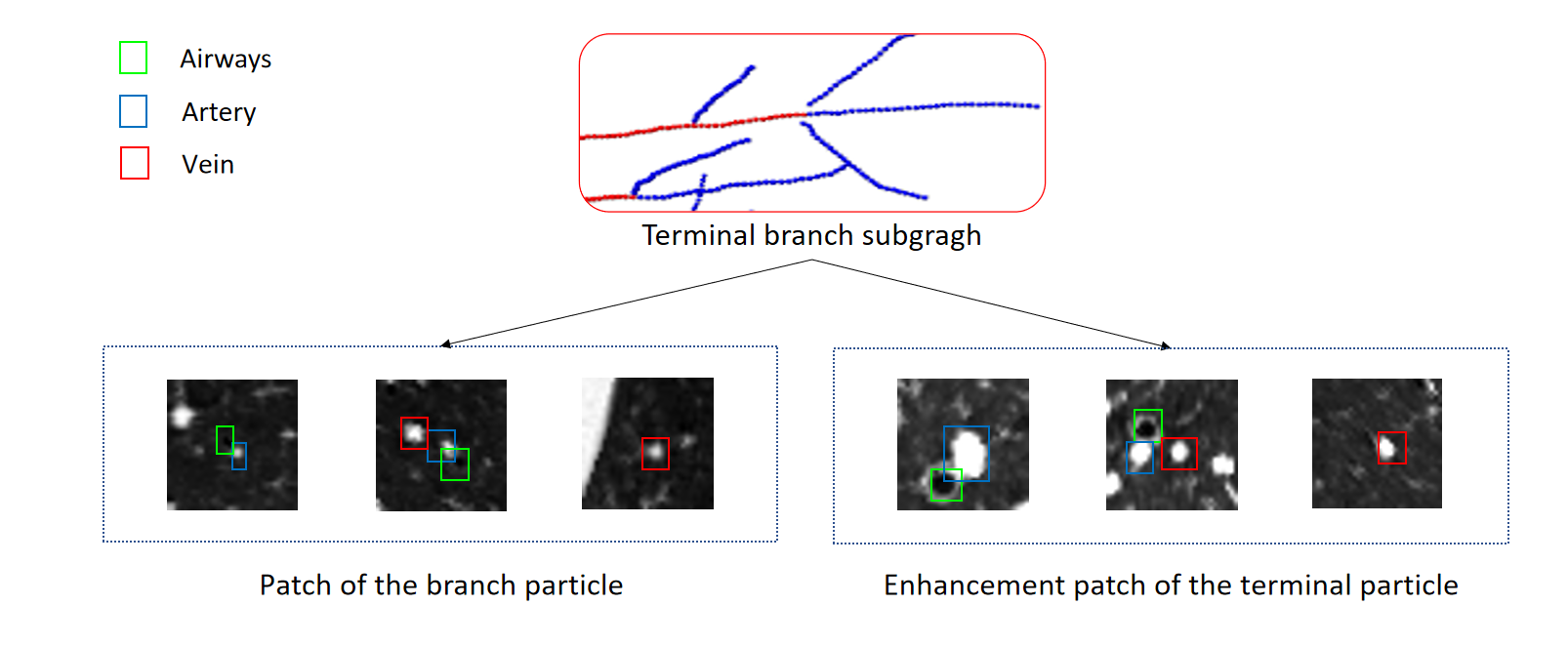}}
\caption{A case of patch extraction process, in which the patch is perpendicular to the direction of CT vessels, and the center is the location of topological vessel particles.}
\label{fig7}
\end{figure*}

\subsubsection{Twin-Pipe network}
{Based on the extracted topology tree, the information of each particle point is used to construct the graph structure we need. Particle points can be divided into three types: terminal points, branch points, and bifurcation points (or attached points, false-positive points). Through the category of particle points and the parent-child node relationship between particles, we redefined a graph T = $\{ X,\varepsilon \}$ composed of nodes X = \{$x_i$\} and edges $\varepsilon$ = \{$\varepsilon_i$\}. And then, in order to train the Twin-Pipe network, we can extract each branch subgragh from graph T, and each branch subgragh is either an artery or a vein.

We propose that the Twin-Pipe network is expected to learn the difference of A/V characteristics caused by different levels and improve the accuracy of the preliminary classification. In this paper, a Twin-Pipe network is designed, in which one pipe network trains the full vessel graph to learn the image information and global connectivity, and the other pipe network specially trains the terminal vessel graph to learn the terminal structure characteristics. Finally, the preliminary classification results are obtained through the mutual correction module. The terminal vessel graph is the set of terminal branches in the full vessel graph, and the extraction process is shown in Fig. 9.

Aiming at the physiological feature of obvious accompanying artery at the end of trachea, for this purpose, the CT original image and vascular enhancement image serve as the input patch of another pipe of Twin-Pipe network, expecting to learn some additional distinguishing features of pulmonary arteries and veins. We perform vascular enhancement and normalization on the original image to enhance the differences between vessels and bronchi. Twin-Pipe network structure design in this paper is shown in the Fig. 10. Fig. 11 shows a case of extraction of terminal vessel particle and patch.}

\subsection{Optimizer and A/V Result}

In order to separate arteries and veins, topological subtrees and topological branches are extracted from topological tree T, and topological subtrees S are roughly extracted from root nodes through uniform distribution of arteries and veins, and subtrees are refined according to tubular features and scale information. The subtree root node in G is extracted, and then traverse each subtree root node down to the terminal points to obtain the corresponding subtree. However, due to the presence of arterial and venous interlacing, it is possible that arterial subtrees contain venous branches, which is inevitable. The topological branch B is extracted from the backtracking path of the end points, where the bifurcation points and terminal points are considered to be the end points to ensure that each branch is either an artery or a vein. The results of a topology subtree and topology branch are shown in Fig. 9.

Though the Twin-Pipe network can learn local, global and connectivity information, spatial inconsistency may still occur during classification. Therefore, after the preliminary classification of Twin-Pipe network, we apply a topology optimizer base on the vascular tree spatial structure to refine the classification. First of all, for the local patch corresponding to each particle, the preliminary probability is given by the Twin-Pipe network. If the probability is greater than 0.5, the node is an artery. otherwise, it is a vein. Then, based on the subtrees extracted in this paper, whether each subtree belongs to an artery or a vein depends on the confidence level of the current subtree. Finally, the branch confidence of each subtree is calculated. When the subtree category and the branch category are inconsistent, and the branch has higher confidence, appropriate pruning is performed to correct the predicted results. The confidence calculation depends on the initial number of predicted arteries and veins on the subtree or branch.

In order to reconstruct pulmonary A/V, base on the refined prediction and scale of center voxels, each voxel in the scale region is labeled with the corresponding prediction of center voxels. It is then fused with vessels near the hilum of the lung for final A/V results.

\section{Experimental methods}

\subsection{Data description}
We trained and evaluated our method on 32 chest CT data from a well-known local first-line 3A hospital, including 16 cases of Siemens noncontrast chest CT scans (Siemens-Non-16), 8 cases of GE noncontrast chest CT scans (GE-Non-8), and 8 cases of GE artery-enhanced chest CT (GE-A-8). The axial images of these CT data were resampled to 512 x 512 sizes, and the slice thickness varies from 0.625 mm to 1 mm. In order to more accurately evaluate the effectiveness of our method, manual annotations of A/V repeatedly confirmed by two experienced chest specialists was used as the reference criteria for evaluation.

In addition, for the purpose of data equalization, 8 cases of the 10 fully annotations of noncontrast chest CT scans from the public CARVE14 dataset were randomly selected to verify the generalization of our method. The CT data were obtained from a Philips scanner with an axial image reconstruction of 512 x 512 size and a thickness of 1.0 mm.

\subsection{Experimental setup details}

Training was performed by running on the Window10 with Intel Core i7-9700 CPU (3.00 GHz), 32 GB RAM, and NVIDIA RTX 2080 GPU with 8 GB of memory. The Twin-Pipe network used the deep learning framework “Tensorflow.Keras” with version 1.12.0. In the training process of the proposed network, we used the SGD optimizer with a momentum of 0.9 and cross-entropy loss function for 150 epochs. The learning rate was 1e-3, and the batch size was 128.

\subsection{Evaluation metrics}

To evaluate the effectiveness of the proposed algorithm, we compare our proposed method with the manual reference standard. For this purpose, we calculate the precision measurement of vessel particles. Then, sensitivity and specificity (considering arteries as the positives) of each method were also computed. In all experiments, the accuracy rate was considered the main evaluation measure, and the sensitivity and specificity were also analyzed to complete the evaluation of the experiment.

\section{Results}

The evaluation structure of the method in the paper is as follows. First, this paper presents the results of our method on the Siemens-Non-16 dataset, as well as the results of the artery-vein segmentation methods in recent years. Second, ablation experiments are performed on the Siemens-Non-16 dataset to validate each component of the method through quantitative results. In addition, generalization experiments are carried out to verify the performance of our method on CT images from GE-Non-8 dataset, GE-A-8 dataset and CARVE14 dataset.

\begin{table*}[t]
    \centering
    \caption{An overview of the results of recent methods.}\label{tbl1}
    \begin{tabular}{ccccc}
        \toprule
        & accuracy & Evaluation parameter & Evaluation method & Dataset\\
        \midrule
        Payer\cite{payer2015automatic}        & 96.3\% & median &  voxel   &  enhanced\\
        Charbonnier\cite{charbonnier2015automatic}  & 89.0\% & median &  voxel   &  noncontrast\\
        Daniel\cite{jimenez2019graph}       & 89.1\% &  mean  & particle &  noncontrast\\
        Nardelli\cite{nardelli2018pulmonary}     & 94.0\% &  mean  & particle &  noncontrast\\
        Zhai\cite{zhai2019linking}         & 77.8\% &  mean  & particle &  enhanced\\
        Yulei Qin\cite{qin2020learning}    & 90.3\% &  mean  &  voxel   &  noncontrast\\
        Our approach & 96.2\% &  mean  & particle &  noncontrast\\
        \bottomrule
    \end{tabular}
\end{table*}

\begin{table*}[t]
    \centering
    \caption{preliminary classification results of a baseline network(Non-local CNN-GCN network) and Twin-Pipe network presented in this paper.}\label{tbl2}
    \begin{tabular}{cccc}
        \toprule
        & baseline network in full vessels & baseline network in terminal vessels & Twin-Pipe network in terminal vessel\\
        \midrule
        Acc.(\%)    &  91.6	& 91.2	& 91.8\\
        Sens.(\%)   &  88.4	& 86.5	& 88.1\\
        Spec.(\%)   &  94.1	& 94.8	& 94.7\\
        \bottomrule
    \end{tabular}
\end{table*}

\subsection{Comparison of recent artery-vein segmentation methods}
Table 1 presents the comparison results of our methods in recent years for pulmonary A/V segmentation. As shown in Table 1, Payer\cite{payer2015automatic} implemented A/V separation on enhanced CT based on two integer programming with an interactive accuracy of 96.3\%. Charbonnier\cite{charbonnier2015automatic} used tree partitioning and peripheral vessel matching to classify arteries and veins with median accuracy of 89.0\% on noncontrast CT. Daniel\cite{jimenez2019graph} demonstrated a significant improvement in noncontrast CT pulmonary vein separation with the graph-cut method. Nardelli et al.\cite{nardelli2018pulmonary} used deep learning to solve parameter optimization and caused the network to learn the difference between arterioles automatically; the overall accuracy of this method reached 94\% in noncontrast CT. Zhai et al.\cite{zhai2019linking} proposed a new network model for end-to-end training; it greatly reduced the complexity of the algorithm and the dependence on parameters. The proposed CNN-GCN method improved lung A/V separation compared with the baseline CNN method. Yulei Qin et al.\cite{qin2020learning} designed end-to-end A/V separation, which was directly realized by the learning tubule-sensitive CNNs. In addition, no presegmentation or postprocessing is designed in the pipeline to avoid the accumulation of errors, and this method reached 90.3\% in public noncontrast CT. Compared with manual reference standard, the proposed artery-vein separation method can achieve an accuracy of 96.2\% on the Siemens-Non-16 dataset.

\subsection{Ablation experiments}
We investigate the effectiveness of the key components of the proposed approach, the Twin-Pipe network, and the topological postprocessing optimization.

\subsubsection{Twin-Pipe network}
{To verify the effectiveness of the proposed Twin-Pipe network, we trained and tested the Siemens-Non-16 dataset, and compared and analyzed the preliminary classification accuracy of terminal vessels using the proposed Twin-Pipe network and baseline Non-local CNN-GCN network. Table 2 shows the classification results of different classification networks in terminal vessels. As shown in Table 2, the classification accuracy, sensitivity, and specificity of the baseline network on the terminal vessels were 91.2\%, 86.5\%, and 94.8\%, respectively. By comparison, the classification accuracy, sensitivity, and specificity of the proposed Twin-Pipe network on the terminal vessels were 91.8\%, 88.1\%, and 94.7\%. That is, the Twin-Pipe network outperforms the baseline network in terms of performance.}

\subsubsection{Topological postprocessing optimizer}
{To prove that the proposed topology optimizer is reasonable, we design different topology strategy optimizer to compare and analyze the experimental results. This approach includes the following: the precision of Twin-Pipe network (based on particle), the precision of topology branch refining (based on branch), the precision of topology subtree refining (based on subtree), and the refinement precision of the proposed topology optimizer. Table 3 shows an overview of the accuracy in all cases of the different topology strategy optimizers and reports on the sensitivity and specificity. The table 3 shows that our method is superior to branch-based and subtree-based topology optimizers, whose accuracies are 96.2\%, 95.3\%, and 93.2\%, respectively.}

\begin{table*}[t]
    \centering
    \caption{An overview of the results obtained compared to other post-processing refinement strategies.}\label{tbl3}
    \begin{tabular}{ccccc}
        \toprule
            & Based on particle	& Based on branch	& Based on subtree	& Our approach\\
        \midrule
            Acc.(\%)    &  91.8	&  95.3	&  93.2	&  96.2\\
            Sens.(\%)   &  89.1	&  93.2	&  92.2	&  94.1\\
            Spec.(\%)   &  94.1	&  96.9	&  93.8	&  97.8\\
        \bottomrule
    \end{tabular}
\end{table*}

\begin{table*}[t]
    \centering
    \caption{An overview of the results obtained under different types of CT scans.}\label{tbl4}
    \begin{tabular}{cccc}
        \toprule
            & the GE-Non-8 dataset	& the GE-A-8 dataset	& CARVE14 dataset \\
        \midrule
            Acc.(\%)    &  93.8	  &  94.8	  &  90.1\\
            Sens.(\%)   &  92.0	  &  91.8	  &  90.7\\
            Spec.(\%)   &  95.5	  &  97.7	  &  89.8\\
        \bottomrule
    \end{tabular}
\end{table*}

\begin{figure}[t]
\centerline{\includegraphics[width=\columnwidth]{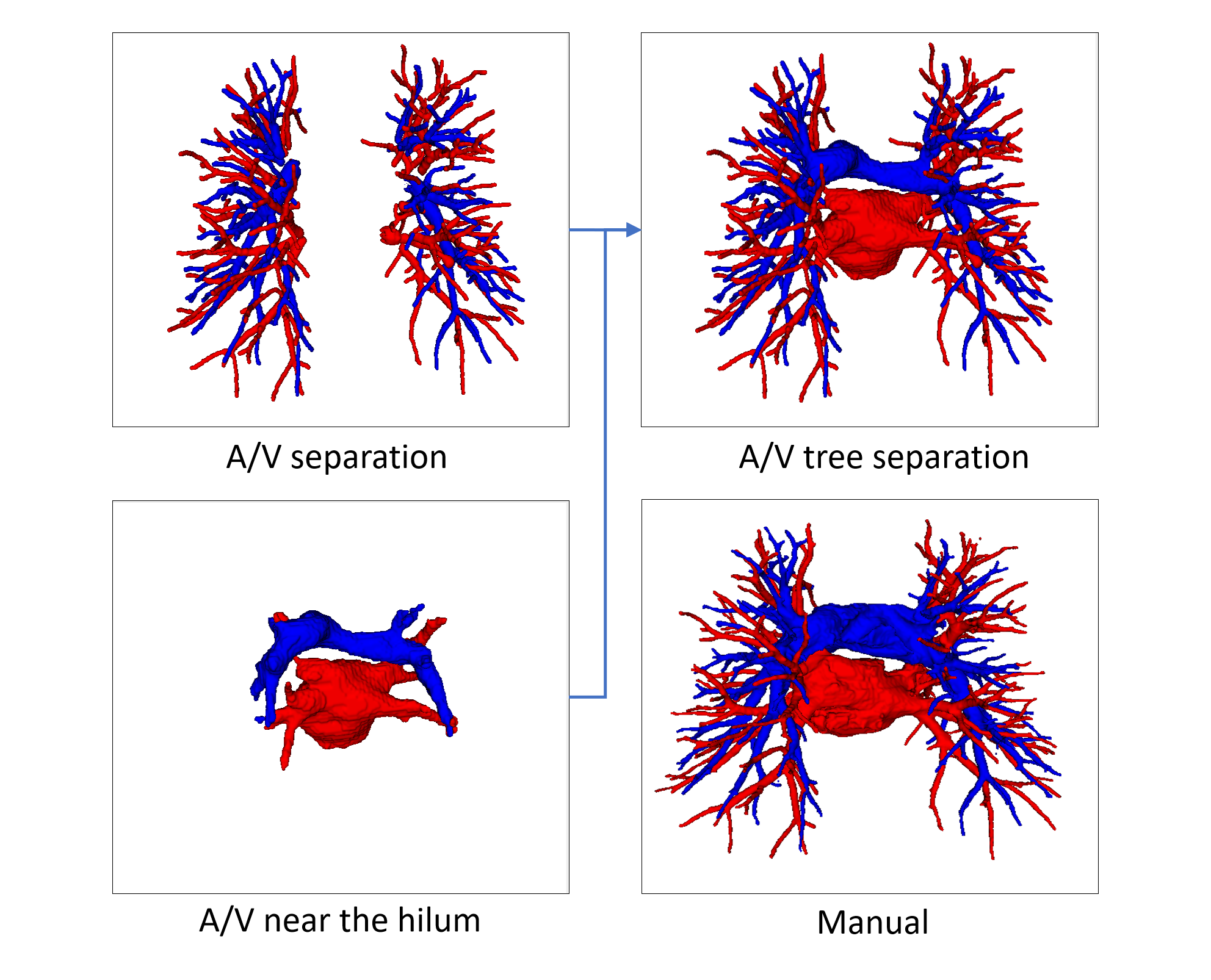}}
\caption{An example of manual separation and the A/V separation presented in this paper.}
\label{fig12}
\end{figure}

\begin{figure}[t]
\centerline{\includegraphics[width=\columnwidth]{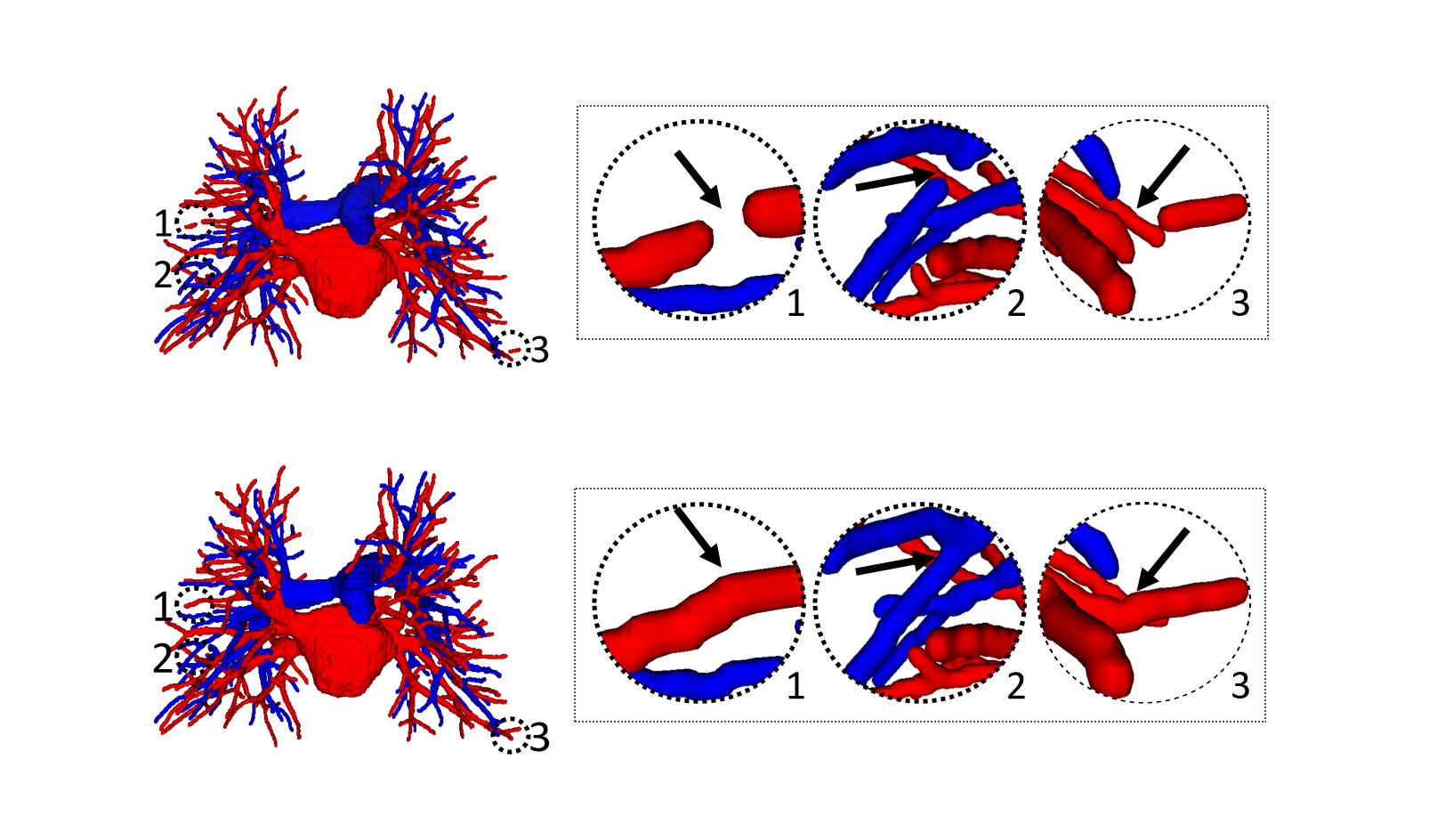}}
\caption{A case of topology reconstruction. The upper row is the result of a scale-space particles reconstruction, and the lower row is the reconstruction result of vascular tree topology extraction. The circular area shows the reconstruction of the local area.}
\label{fig13}
\end{figure}

\subsection{Generalization experiments}

{In order to verify the generalization of the proposed method. We performed validation on three other different data sets: the GE-Non-8 dataset from another device GE, the GE-A-8 dataset from another mode, and publicly available the CARVE14 dataset. Table 4 shows the summary of case accuracy under different types of CT scans, from which it can be seen that the accuracy of this method on the GE-Non-8 dataset, the GE-A-8 dataset and the CARVE14 dataset were 93.8\%, 94.8\% and 90.1\%, respectively.}

\section{Discussion and Conclusion}

We propose a new method for automatic separation of pulmonary arteries and veins. We extract vascular skeleton through the vascular tree topology method, obtain the preliminary classification results after Twin-Pipe network training, and finally use the topology structure information for postprocessing to refine the classification results. This method is suitable for noncontrast CT scans that lack vascular edge information. The experimental results in Table 1 indicate that the average accuracy of the proposed method can reach 96.2\% compared with that of manual reference annotation. The experimental results in Table 4 shows that our method is also applicable to CT image from different devices and different modes. Experimental results show that this method has good performance.

Our evaluation method for A/V separation is based on topological particles rather than voxel classification. This finding is mainly due to the difficulty of obtaining accurate annotations of pulmonary vessels based on voxel levels, which are also indistinguishable by doctors. In addition, in clinical practice, doctors focus more on the structural branching direction of vessels, consistent with the evaluation system based on topological particles. A/V separation can provide effective information for doctors to make surgical plans and perform surgical navigation, and proximal vascular extension can help doctors to locate the vessel more quickly. Therefore, the final A/V separation results include vessels close to the hilum of the lung, as shown in Fig. 12. Our evaluation system for the method of A/V separation do not include vessels near the hilum of the lung mainly because the vessels near the hilum are abnormally large and nontubular, and the vascular topology could not be extracted.

\begin{figure*}
\centerline{\includegraphics[width=1\linewidth]{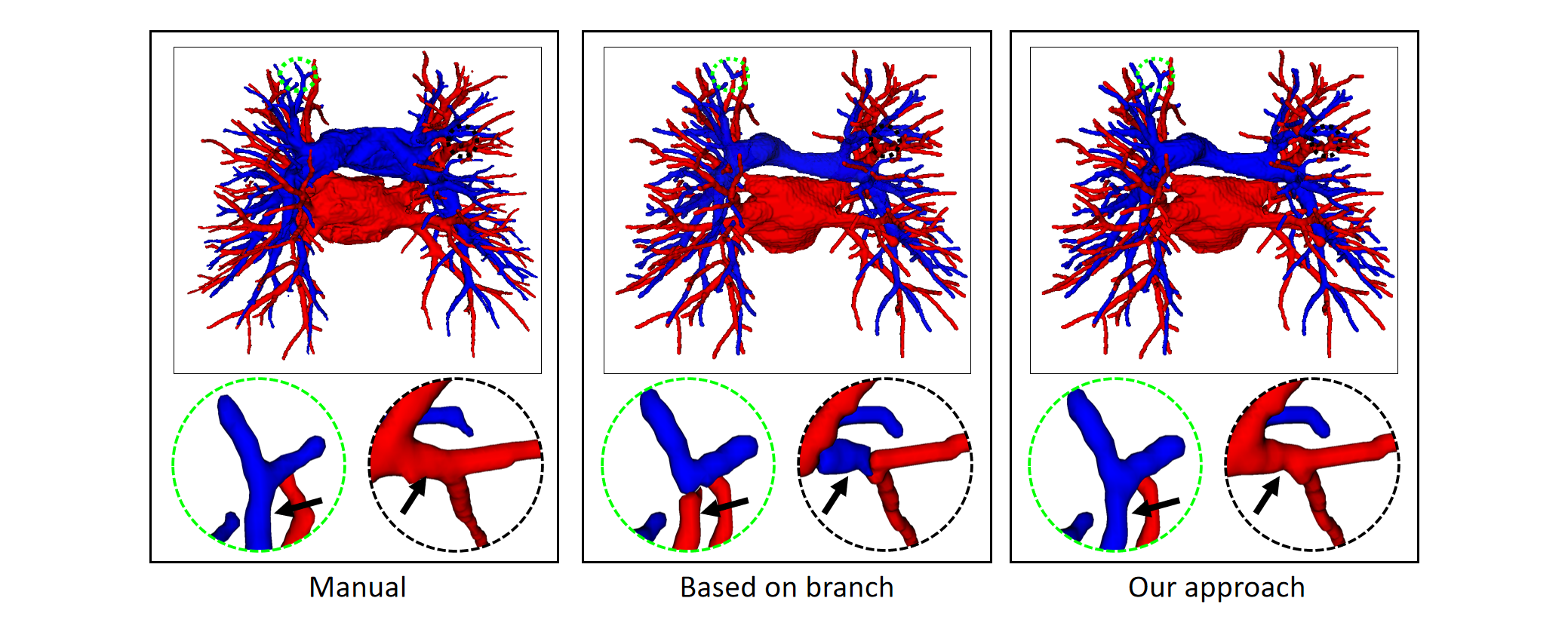}}
\caption{In an example of A/V separation results, the three columns are manually annotated, based on branch optimization separation, and our method (from left to right), the circular area highlights local classification results. Due to the complex and changeable structure of the vascular tree, the local circular display position has been translated, rotated and enlarged from the original image position, which is different from the original image position to some extent.}
\label{fig14}
\end{figure*}

\begin{figure*}
\centerline{\includegraphics[width=1\linewidth]{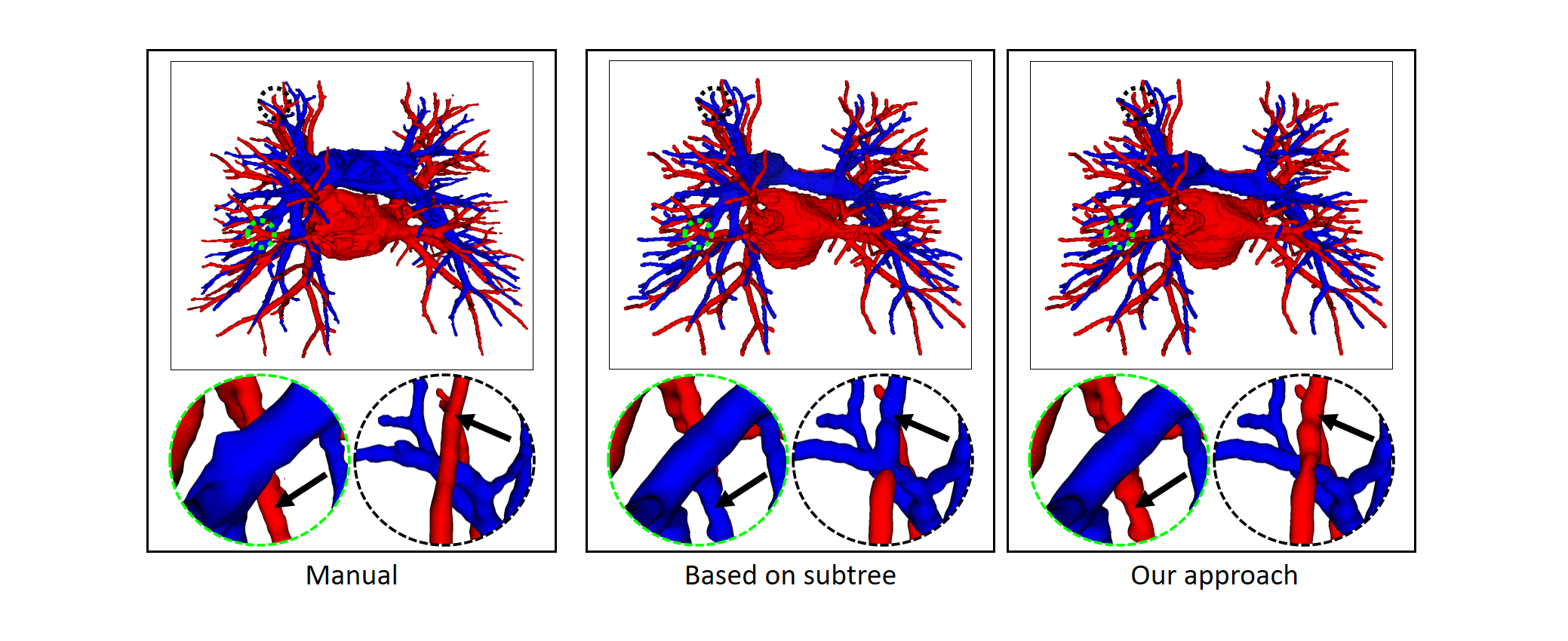}}
\caption{In an example of A/V separation results, the three columns are manually annotated, based on subtree optimization separation, and our method (from left to right), the circular area highlights local classification results. Due to the complex and changeable structure of the vascular tree, the local circular display position has been translated, rotated and enlarged from the original image position, which is different from the original image position to some extent.}
\label{fig15}
\end{figure*}

\begin{figure}[t]
\centerline{\includegraphics[width=\columnwidth]{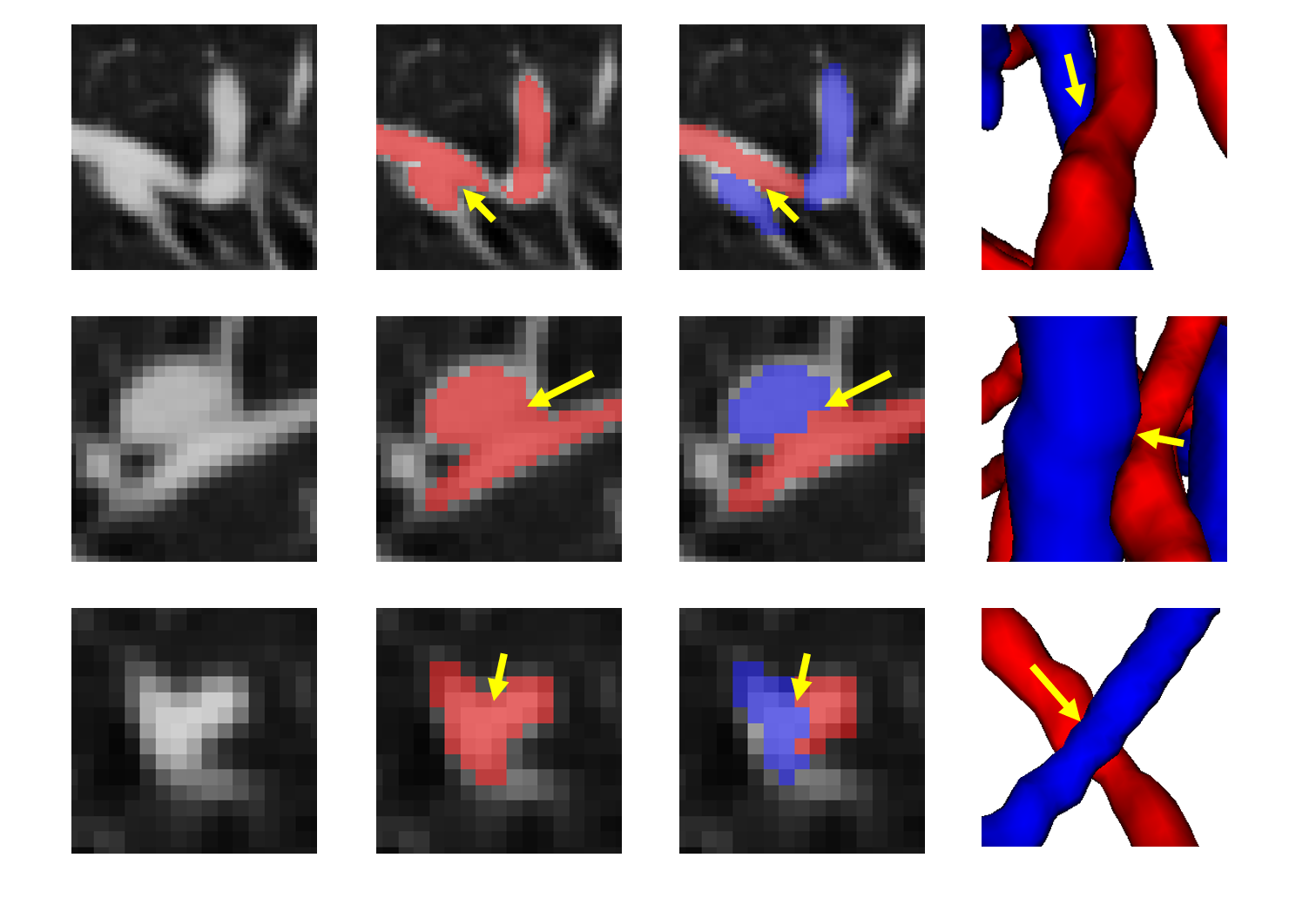}}
\caption{The results of separation of arterial and venous intersections are shown in this paper. Each row represents a different location where arteries and veins intersect. The number of columns respectively represents the location of the intersecting arteries and veins in the original CT image, the results of vascular segmentation, the results of the A/V separation method presented in this paper, and the results of the 3D reconstruction of this location. The blue areas represent the arteries and the red veins. The results of 3D reconstruction show that this method can effectively separate artery and vein intersections.}
\label{fig16}
\end{figure}

An objective inter-work comparison could be difficult due to the difference in medical data sets and implementation. Our experiments are mainly conducted on our own CT data.

First, we use the vascular tree topology method to extract vascular topology for the training of classification network. The current topology extraction algorithms have their own applicability, and we propose a vascular tree topology method, which fully uses their advantages to solve the problem of topological fracture caused by particle loss. As shown in Figs. 7 and 13, a complete vascular skeleton structure not only affects the vascular classification evaluation system, but also plays a crucial role in arterial/vein vessel reconstruction.

Second, we design a Twin-Pipe network. As shown in Table 2, the classification accuracy of terminal vessels in the baseline network is lower than that of the entire vessels. The experimental results show that the error frequency is higher in the terminal vessels in the baseline network. This finding is mainly due to the fact that small vessels are more visible as bright tubes than bronchi of the same size. Moreover, the radius of the bronchioles is smaller than that of the accompanying arteries, and the walls of the bronchioles are susceptible to partial volume. The experimental results are shown in Table 2. In the terminal vessels, the sensitivity of the Twin-Pipe network is higher than that of the baseline network. That is, it can effectively improve the accuracy of terminal artery classification. The experiment proves that the Non-local CNN-GCN network can effectively combine local information, global image information, and graph connection information. Thus, differences in A/V characteristics can be automatically learned, eliminating the need for complex parameter tuning. Moreover, the Twin-Pipe network can effectively learn the difference of A/V characteristics at different levels and the information that the terminal  bronchus is closely associated with the artery. It is superior to the baseline Non-local CNN-GCN network and does not need to depend on the results of trachea segmentation.

Finally, the topology optimizer extracts the topology subtree and topology branch refinement results by the method in Section 2.3. The results of a topology subtree and topology branch are shown in Fig. 9. Then, we use topology subtrees and topology branches for postprocessing. As shown in Table 3, the proposed topology optimizer is superior to the subtree-based and branch-based topology optimization in accuracy. Fig. 14 and 15 show the reconstructed results of different topology strategy optimizers. The results in Fig. 14 show that when the number of points on the branch is small, and the branch-based topology optimization method is prone to predict errors. This finding is mainly because the branch-based refinement strategy focuses on the relationships within the branches and ignores the topological relationships between branches. As shown in Fig. 15, topology optimization based on subtree strategy is prone to prediction errors in the case of arterial-vein intersection. This finding is mainly because the arterial/venous intersection point in the extraction process of topological subtree is easily mistaken for the bifurcation point of arterial subtrees due to the complex structure of the vascular tree, arteries and veins interweave, and finally the venous branches are classified as arterial subtrees, resulting in the classification error. 

Fig. 16 shows the final separation result of our method at the intersections of the arteries and veins. The proposed topology optimizer considers the interbranch and intrabranch relationships. The extracted topological subtree is used to maintain the spatial connectivity of topological particles. The branch confidence calculated using topological branches is used to correct the topological subtree. To some extent, our method can solve the misclassification problem caused by interlaced arteries and veins. 

The results in Table 4 show that the accuracy of this method was 93.8\% on the GE-Non-8 dataset from another device GE. Similarly, under the influence of dynamic changes of arterial enhancement effect with time and contrast agent dose, the characteristics are varied and the difficulty of A/V separation is increased, the accuracy of this method the GE-A-8 dataset from another mode can still reach 94.8\%. This finding proved that the proposed method still maintained a high classification accuracy under different devices and different modes. In view of the similarity between arterial enhanced CT and venous enhanced CT, this paper takes the arterial enhanced CT experiment as an example.

In addition,, we verify the proposed method on the public CARVE14 dataset. Some voxels were annotated as uncertain vessels due to the difficulty in annotating this data set, and pulmonary vessels near the hilum were not annotated. The continuous vascular tree of annotation cannot be guaranteed, and a certain gap from clinical application is observed. Finally, some key components proposed are used for simple prediction, and the classification accuracy of 90.1\% is obtained.
 
This method has made some achievements in A/V separation, but there are still some unavoidable shortcomings. On the one hand, although it is possible to classify the arterial veins at the intersections, the scale of the intersections after reconstruction is abnormal and often resulting in fracture or abnormal expansion. This condition is due to the fact that topological particles at nontubular intersections of arteries and veins cannot be accurately extracted during topology construction, thereby leading to topological particle scale mutation or deletion. On the other hand, our postprocessing strategy refined the preliminary prediction results based on the Twin-Pipe network to ensure the accuracy of the preliminary prediction as much as possible. We believe that increasing the amount of training data or adding the enhanced CT scan data to the training set would surely achieve better results. In addition, given that the proposed method is based on topological particle reconstruction of A/V classification, some errors are found between the reconstructed vessel size and the actual situation. In order to apply it to clinical practice, in the future, we can further improve the classification results and performance by optimizing the topology extraction algorithm and detecting the intersection situation, as well as increase the clinical applicability through semi-automated graphical user interface interaction operation.

\section*{Acknowledgements}
 This project was supported by the Natural Science Foundation of Fujian Province, China (Grant No. 2020J01472) and Provincial Science and Technology Leading Project (Grant No. 2018Y0032).

\bibliographystyle{model1-num-names}

\bibliography{AV_bib}

\begin{thebibliography}{18}
\expandafter\ifx\csname natexlab\endcsname\relax\def\natexlab#1{#1}\fi
\providecommand{\url}[1]{\texttt{#1}}
\providecommand{\href}[2]{#2}
\providecommand{\path}[1]{#1}
\providecommand{\DOIprefix}{doi:}
\providecommand{\ArXivprefix}{arXiv:}
\providecommand{\URLprefix}{URL: }
\providecommand{\Pubmedprefix}{pmid:}
\providecommand{\doi}[1]{\href{http://dx.doi.org/#1}{\path{#1}}}
\providecommand{\Pubmed}[1]{\href{pmid:#1}{\path{#1}}}
\providecommand{\bibinfo}[2]{#2}
\ifx\xfnm\relax \def\xfnm[#1]{\unskip,\space#1}\fi
\bibitem[{Zheng et~al.(2019)Zheng, Sun, Zhang, Zeng, Zou, Chen, Gu, Wei, and
  He}]{zheng2019report}
\bibinfo{author}{R.~t. Zheng}, \bibinfo{author}{K.~t. Sun},
  \bibinfo{author}{S.~t. Zhang}, \bibinfo{author}{H.~t. Zeng},
  \bibinfo{author}{X.~t. Zou}, \bibinfo{author}{R.~Chen},
  \bibinfo{author}{X.~t. Gu}, \bibinfo{author}{W.~t. Wei},
  \bibinfo{author}{J.~He},
\newblock \bibinfo{title}{Report of cancer epidemiology in china, 2015},
\newblock \bibinfo{journal}{Zhonghua zhong liu za zhi [Chinese journal of
  oncology]} \bibinfo{volume}{41} (\bibinfo{year}{2019})
  \bibinfo{pages}{19--28}.
\bibitem[{Tozaki et~al.(2001)Tozaki, Kawata, Niki, and
  Ohmatsu}]{tozaki2001extraction}
\bibinfo{author}{T.~Tozaki}, \bibinfo{author}{Y.~Kawata},
  \bibinfo{author}{N.~Niki}, \bibinfo{author}{H.~Ohmatsu},
\newblock \bibinfo{title}{Extraction and classification of pulmonary organs
  based on thoracic 3d ct images},
\newblock \bibinfo{journal}{Systems and Computers in Japan}
  \bibinfo{volume}{32} (\bibinfo{year}{2001}) \bibinfo{pages}{42--53}.
\bibitem[{Nakamura et~al.(2005)Nakamura, Mekada, Ide, Murase, Hasegawa,
  Toriwaki, and Otsuji}]{nakamura2005pulmonary}
\bibinfo{author}{S.~Nakamura}, \bibinfo{author}{Y.~Mekada},
  \bibinfo{author}{I.~Ide}, \bibinfo{author}{H.~Murase},
  \bibinfo{author}{J.~Hasegawa}, \bibinfo{author}{J.~Toriwaki},
  \bibinfo{author}{H.~Otsuji},
\newblock \bibinfo{title}{Pulmonary artery and vein classification method using
  spatial arrangement features from x-ray ct image},
\newblock in: \bibinfo{booktitle}{International Congress Series}, volume
  \bibinfo{volume}{1281}, \bibinfo{organization}{Citeseer},
  \bibinfo{year}{2005}, pp. \bibinfo{pages}{1403--1403}.
\bibitem[{B{\"u}low et~al.(2005)B{\"u}low, Wiemker, Blaffert, Lorenz, and
  Renisch}]{bulow2005automatic}
\bibinfo{author}{T.~B{\"u}low}, \bibinfo{author}{R.~Wiemker},
  \bibinfo{author}{T.~Blaffert}, \bibinfo{author}{C.~Lorenz},
  \bibinfo{author}{S.~Renisch},
\newblock \bibinfo{title}{Automatic extraction of the pulmonary artery tree
  from multi-slice ct data},
\newblock in: \bibinfo{booktitle}{Medical Imaging 2005: Physiology, Function,
  and Structure from Medical Images}, volume \bibinfo{volume}{5746},
  \bibinfo{organization}{International Society for Optics and Photonics},
  \bibinfo{year}{2005}, pp. \bibinfo{pages}{730--740}.
\bibitem[{Saha et~al.(2010)Saha, Gao, Alford, Sonka, and
  Hoffman}]{saha2010topomorphologic}
\bibinfo{author}{P.~K. Saha}, \bibinfo{author}{Z.~Gao}, \bibinfo{author}{S.~K.
  Alford}, \bibinfo{author}{M.~Sonka}, \bibinfo{author}{E.~A. Hoffman},
\newblock \bibinfo{title}{Topomorphologic separation of fused isointensity
  objects via multiscale opening: Separating arteries and veins in 3-d
  pulmonary ct},
\newblock \bibinfo{journal}{IEEE transactions on medical imaging}
  \bibinfo{volume}{29} (\bibinfo{year}{2010}) \bibinfo{pages}{840--851}.
\bibitem[{Wala et~al.(2011)Wala, Fotin, Lee, Jirapatnakul, Biancardi, and
  Reeves}]{wala2011automated}
\bibinfo{author}{J.~Wala}, \bibinfo{author}{S.~Fotin},
  \bibinfo{author}{J.~Lee}, \bibinfo{author}{A.~Jirapatnakul},
  \bibinfo{author}{A.~Biancardi}, \bibinfo{author}{A.~Reeves},
\newblock \bibinfo{title}{Automated segmentation of the pulmonary arteries in
  low-dose ct by vessel tracking},
\newblock \bibinfo{journal}{arXiv preprint arXiv:1106.5460}
  (\bibinfo{year}{2011}).
\bibitem[{Kitamura et~al.(2016)Kitamura, Li, Ito, and
  Ishikawa}]{kitamura2016data}
\bibinfo{author}{Y.~Kitamura}, \bibinfo{author}{Y.~Li},
  \bibinfo{author}{W.~Ito}, \bibinfo{author}{H.~Ishikawa},
\newblock \bibinfo{title}{Data-dependent higher-order clique selection for
  artery--vein segmentation by energy minimization},
\newblock \bibinfo{journal}{International Journal of Computer Vision}
  \bibinfo{volume}{117} (\bibinfo{year}{2016}) \bibinfo{pages}{142--158}.
\bibitem[{Park et~al.(2013)Park, Min~Lee, Kim, Beom~Seo, and
  Shin}]{park2013automatic}
\bibinfo{author}{S.~Park}, \bibinfo{author}{S.~Min~Lee},
  \bibinfo{author}{N.~Kim}, \bibinfo{author}{J.~Beom~Seo},
  \bibinfo{author}{H.~Shin},
\newblock \bibinfo{title}{Automatic reconstruction of the arterial and venous
  trees on volumetric chest ct},
\newblock \bibinfo{journal}{Medical physics} \bibinfo{volume}{40}
  (\bibinfo{year}{2013}) \bibinfo{pages}{071906}.
\bibitem[{Payer et~al.(2015)Payer, Pienn, B{\'a}lint, Olschewski, Olschewski,
  and Urschler}]{payer2015automatic}
\bibinfo{author}{C.~Payer}, \bibinfo{author}{M.~Pienn},
  \bibinfo{author}{Z.~B{\'a}lint}, \bibinfo{author}{A.~Olschewski},
  \bibinfo{author}{H.~Olschewski}, \bibinfo{author}{M.~Urschler},
\newblock \bibinfo{title}{Automatic artery-vein separation from thoracic ct
  images using integer programming},
\newblock in: \bibinfo{booktitle}{International Conference on Medical Image
  Computing and Computer-Assisted Intervention},
  \bibinfo{organization}{Springer}, \bibinfo{year}{2015}, pp.
  \bibinfo{pages}{36--43}.
\bibitem[{Charbonnier et~al.(2015)Charbonnier, Brink, Ciompi, Scholten,
  Schaefer-Prokop, and Van~Rikxoort}]{charbonnier2015automatic}
\bibinfo{author}{J.-P. Charbonnier}, \bibinfo{author}{M.~Brink},
  \bibinfo{author}{F.~Ciompi}, \bibinfo{author}{E.~T. Scholten},
  \bibinfo{author}{C.~M. Schaefer-Prokop}, \bibinfo{author}{E.~M.
  Van~Rikxoort},
\newblock \bibinfo{title}{Automatic pulmonary artery-vein separation and
  classification in computed tomography using tree partitioning and peripheral
  vessel matching},
\newblock \bibinfo{journal}{IEEE transactions on medical imaging}
  \bibinfo{volume}{35} (\bibinfo{year}{2015}) \bibinfo{pages}{882--892}.
\bibitem[{Jimenez-Carretero et~al.(2019)Jimenez-Carretero, Bermejo-Pel{\'a}ez,
  Nardelli, Fraga, Fraile, Est{\'e}par, and Ledesma-Carbayo}]{jimenez2019graph}
\bibinfo{author}{D.~Jimenez-Carretero},
  \bibinfo{author}{D.~Bermejo-Pel{\'a}ez}, \bibinfo{author}{P.~Nardelli},
  \bibinfo{author}{P.~Fraga}, \bibinfo{author}{E.~Fraile},
  \bibinfo{author}{R.~S.~J. Est{\'e}par}, \bibinfo{author}{M.~J.
  Ledesma-Carbayo},
\newblock \bibinfo{title}{A graph-cut approach for pulmonary artery-vein
  segmentation in noncontrast ct images},
\newblock \bibinfo{journal}{Medical image analysis} \bibinfo{volume}{52}
  (\bibinfo{year}{2019}) \bibinfo{pages}{144--159}.
\bibitem[{Nardelli et~al.(2018)Nardelli, Jimenez-Carretero, Bermejo-Pelaez,
  Washko, Rahaghi, Ledesma-Carbayo, and Est{\'e}par}]{nardelli2018pulmonary}
\bibinfo{author}{P.~Nardelli}, \bibinfo{author}{D.~Jimenez-Carretero},
  \bibinfo{author}{D.~Bermejo-Pelaez}, \bibinfo{author}{G.~R. Washko},
  \bibinfo{author}{F.~N. Rahaghi}, \bibinfo{author}{M.~J. Ledesma-Carbayo},
  \bibinfo{author}{R.~S.~J. Est{\'e}par},
\newblock \bibinfo{title}{Pulmonary artery--vein classification in ct images
  using deep learning},
\newblock \bibinfo{journal}{IEEE transactions on medical imaging}
  \bibinfo{volume}{37} (\bibinfo{year}{2018}) \bibinfo{pages}{2428--2440}.
\bibitem[{Zhai et~al.(2019)Zhai, Staring, Zhou, Xie, Xiao, Bakker, Kroft,
  Lelieveldt, Boon, Klok et~al.}]{zhai2019linking}
\bibinfo{author}{Z.~Zhai}, \bibinfo{author}{M.~Staring},
  \bibinfo{author}{X.~Zhou}, \bibinfo{author}{Q.~Xie},
  \bibinfo{author}{X.~Xiao}, \bibinfo{author}{M.~E. Bakker},
  \bibinfo{author}{L.~J. Kroft}, \bibinfo{author}{B.~P. Lelieveldt},
  \bibinfo{author}{G.~J. Boon}, \bibinfo{author}{F.~A. Klok}, et~al.,
\newblock \bibinfo{title}{Linking convolutional neural networks with graph
  convolutional networks: application in pulmonary artery-vein separation},
\newblock in: \bibinfo{booktitle}{International Workshop on Graph Learning in
  Medical Imaging}, \bibinfo{organization}{Springer}, \bibinfo{year}{2019}, pp.
  \bibinfo{pages}{36--43}.
\bibitem[{Qin et~al.(2020)Qin, Zheng, Gu, Huang, Yang, Wang, Yao, Zhu, and
  Yang}]{qin2020learning}
\bibinfo{author}{Y.~Qin}, \bibinfo{author}{H.~Zheng}, \bibinfo{author}{Y.~Gu},
  \bibinfo{author}{X.~Huang}, \bibinfo{author}{J.~Yang},
  \bibinfo{author}{L.~Wang}, \bibinfo{author}{F.~Yao}, \bibinfo{author}{Y.-M.
  Zhu}, \bibinfo{author}{G.-Z. Yang},
\newblock \bibinfo{title}{Learning tubule-sensitive cnns for pulmonary airway
  and artery-vein segmentation in ct},
\newblock \bibinfo{journal}{arXiv preprint arXiv:2012.05767}
  (\bibinfo{year}{2020}).
\bibitem[{Cui et~al.(2019)Cui, Liu, and Huang}]{cui2019pulmonary}
\bibinfo{author}{H.~Cui}, \bibinfo{author}{X.~Liu}, \bibinfo{author}{N.~Huang},
\newblock \bibinfo{title}{Pulmonary vessel segmentation based on orthogonal
  fused u-net++ of chest ct images},
\newblock in: \bibinfo{booktitle}{International Conference on Medical Image
  Computing and Computer-Assisted Intervention},
  \bibinfo{organization}{Springer}, \bibinfo{year}{2019}, pp.
  \bibinfo{pages}{293--300}.
\bibitem[{Saha et~al.(2016)Saha, Borgefors, and di~Baja}]{saha2016survey}
\bibinfo{author}{P.~K. Saha}, \bibinfo{author}{G.~Borgefors},
  \bibinfo{author}{G.~S. di~Baja},
\newblock \bibinfo{title}{A survey on skeletonization algorithms and their
  applications},
\newblock \bibinfo{journal}{Pattern recognition letters} \bibinfo{volume}{76}
  (\bibinfo{year}{2016}) \bibinfo{pages}{3--12}.
\bibitem[{Liu et~al.(2018)Liu, Zhang, Song, Peng, and Cai}]{liu2018automated}
\bibinfo{author}{S.~Liu}, \bibinfo{author}{D.~Zhang},
  \bibinfo{author}{Y.~Song}, \bibinfo{author}{H.~Peng},
  \bibinfo{author}{W.~Cai},
\newblock \bibinfo{title}{Automated 3-d neuron tracing with precise branch
  erasing and confidence controlled back tracking},
\newblock \bibinfo{journal}{IEEE transactions on medical imaging}
  \bibinfo{volume}{37} (\bibinfo{year}{2018}) \bibinfo{pages}{2441--2452}.
\bibitem[{Est{\'e}par et~al.(2012)Est{\'e}par, Ross, Russian, Schultz, Washko,
  and Kindlmann}]{estepar2012computational}
\bibinfo{author}{R.~S.~J. Est{\'e}par}, \bibinfo{author}{J.~C. Ross},
  \bibinfo{author}{K.~Russian}, \bibinfo{author}{T.~Schultz},
  \bibinfo{author}{G.~R. Washko}, \bibinfo{author}{G.~L. Kindlmann},
\newblock \bibinfo{title}{Computational vascular morphometry for the assessment
  of pulmonary vascular disease based on scale-space particles},
\newblock in: \bibinfo{booktitle}{2012 9th IEEE International Symposium on
  Biomedical Imaging (ISBI)}, \bibinfo{organization}{IEEE},
  \bibinfo{year}{2012}, pp. \bibinfo{pages}{1479--1482}.

\end{thebibliography}

\end{document}